\documentclass[12pt, a4paper]{iopart}
\usepackage{graphicx}
\usepackage{subfig}
\usepackage{iopams}
\usepackage{setstack}
\usepackage{cite}
\usepackage{psfrag}
\usepackage{wasysym}
\usepackage{color}
\newcommand*{\pbar}{\ensuremath{\overline{\mathrm{p}}}}
\newcommand*{\pbHe}{\ensuremath{\overline{\mathrm{p}}\mathrm{He}^+}}

\begin{document}

\title{Collision Induced Relaxations within the Antiprotonic Helium Hyperfine
Structure}
\author{T Pask$^{1}$, D~Barna$^{2,3}$, A~Dax$^{2}$, S~Friedreich$^{1}$,
R~S~Hayano$^2$, M~Hori$^{2,4}$, D Horv\'ath$^{3,5}$, B~Juh\'asz$^1$,
C~Malbrunot$^1$\footnote{Present address: TRIUMF, 4004 Wesbrook Mall, Vancouver,
BC, V6T 2A3, Canada.}, O~Massiczek$^1$, N Ono$^2$, A~S\'ot\'er$^3$ and
E~Widmann$^1$}

\address{$^1$ Stefan Meyer Institute for Subatomic Physics, Austrian Academy of
Sciences, Boltzmanngasse 3, A-1090 Vienna, Austria.} \ead{thomas.pask@cern.ch}

\address{$^2$ Department of Physics, University of Tokyo, 7-3-1 Hongo,
Bunkyo-ku, Tokyo 113-0033, Japan}

\address{$^3$ KFKI Research Institute for Particle and Nuclear Physics, H-1525
Budapest, PO Box 49, Hungary}

\address{$^4$ Max-Planck-Institut f\"{u}r Quantenoptik, Hans-Kopfermann-Strasse
1, D-85748 Garching, Germany.}

\address{$^5$ Institute of Nuclear Research of the Hungarian Academy of
Sciences, H-4001 Debrecen, PO Box 51, Hungary}

\begin{abstract}
We report the first measurements of the inelastic spin exchange collision rate
between the Hyperfine (HF) levels of antiprotonic helium ($\pbHe$).  We measure
the time dependent evolution of the $(37, \, 35)$ substates to obtain an
inelastic collision rate which qualitatively agrees with recent
theoretical calculations.  We evaluate these results by using the obtained rate
as a parameter in a rigorous simulation which we then compare to  to previously
measured data. We find that our measurement slightly underestimates the
collision rate and therefore conclude that the actual value most probably falls
within the upper, rather than lower, limit of the error.

\end{abstract}

\submitto{\jpb}

\maketitle

\section{Introduction}

\begin{figure}
\subfloat[]{
\label{fig:subfig:cascade}
\includegraphics[scale=0.5]{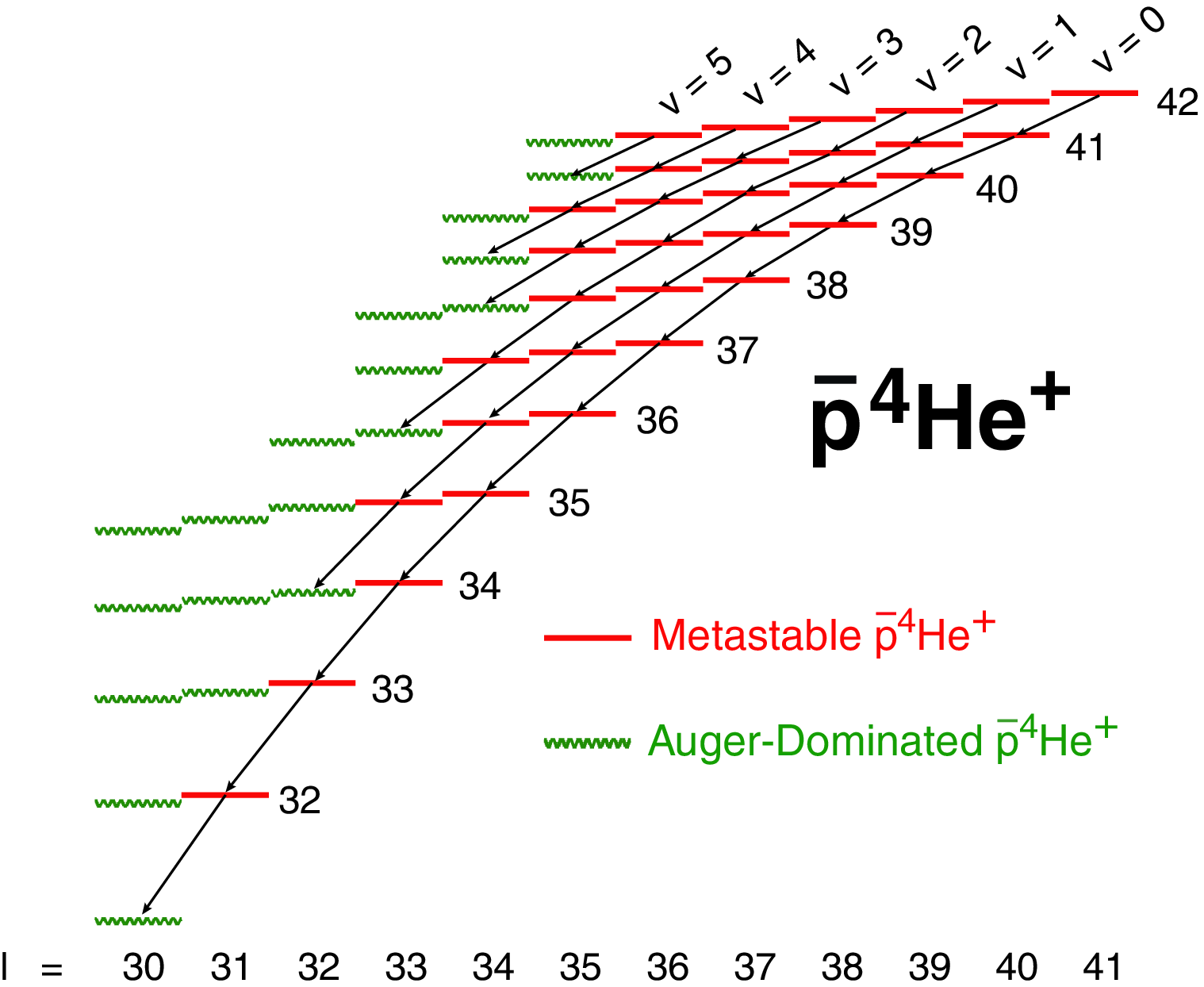}}
\subfloat[]{
\label{fig:subfig:HFS}
\includegraphics[scale=0.35]{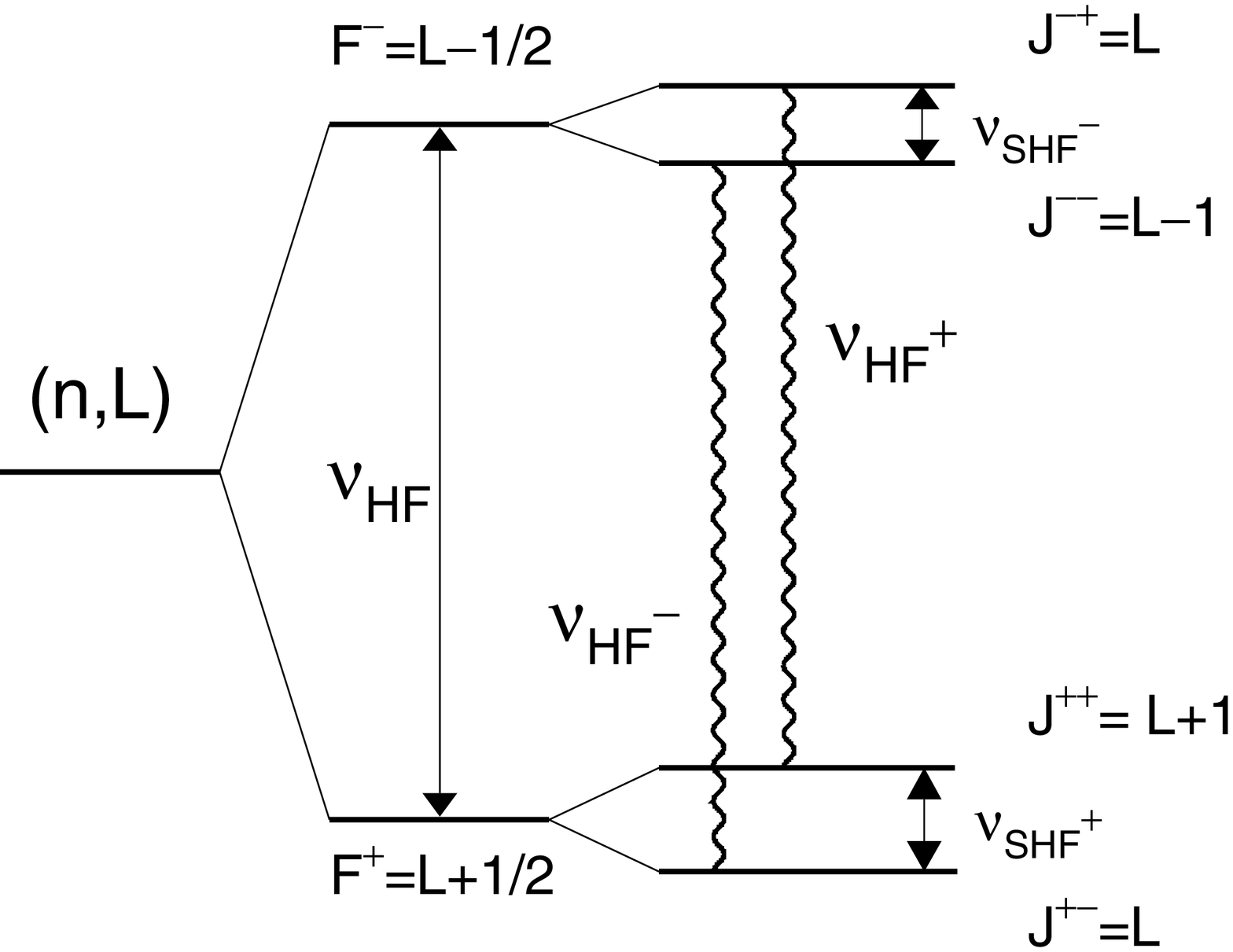}}
\caption{(a) Level diagram of $\pbar^4 \mathrm{He}^+$ where the arrows indicate
the radiative cascade towards the nucleus. (b) Hyperfine splitting of an ($n, \,
l$) state  of $\pbar^4$He$^+$. The wavy lines denote allowed M1 transitions that
can be induced by an oscillating magnetic field. From~\cite{Hayano:2007}.}
\label{fig:HFS}
\end{figure}

Antiprotonic helium $\pbHe$ is formed when an antiproton $\pbar$ interacts with
a helium atom at or below the ionization energy ($\sim
25$~eV)~\cite{Iwasaki:91}.  The $\pbar$ can become captured so that it precesses
around the helium nucleus He$^{++}$~\cite{Yamazaki:93,Yamazaki:02,Hayano:2007}. 
When this happens, one of the electrons e$^-$ is ejected.  Because of its mass,
the $\pbar$ is most likely to occupy an orbit with principle quantum number $n =
n_0 \equiv \sqrt{M^*/m_{\mathrm{e}}} \sim 38$~\cite{Condo:64}, where $M^*$ is
the reduced mass of the antiproton-helium nucleus system and $m_{\mathrm{e}}$ is
the electron mass.  It precesses in a semi-classical orbit while the electron
remains in a 1s quantum mechanical cloud.

Because of their overlap with the nucleus, the majority of captured antiprotons
annihilate within picoseconds with one of the nucleons in the He$^{++}$
nucleus~\cite{Yamazaki:02}.  However, approximately 3\% occupy metastable
states, so called circular states in the region of $n = 32$-40 and vibrational
quantum number $v = 0$-3 (where $v = n - l - 1$ and $l$ is the angular momentum
quantum number).  Since the neutral system retains one electron, it is protected
from external atoms by the Pauli exclusion principle~\cite{Yamazaki:02}. 
Additionally, the presence of the electron removes the $l$ degeneracy for the
same
$n$, therefore protecting it against Stark mixing.  The Auger decay of the
remaining electron is suppressed by the large ionization energy compared to the
$n \rightarrow n - 1$ level spacing of $\sim 2$~eV.  Thus only one decay channel
remains and the antiprotons in these states undergo a radiative cascade through
($n, \, l) \rightarrow (n-1, \, l-1)$ states, each with lifetimes in the order
$\sim 1.5 \, \mu$s, see figure~\ref{fig:subfig:cascade}.

A \emph{hyperfine (HF)} splitting~\cite{Yamazaki:02}, caused by the interaction
of the e$^-$ spin $S_{\mathrm{e}}$ with the $\pbar$ orbital angular
momentum $L$, results in a doublet structure of the order $\nu_{\mathrm{HF}} =
10$-15~GHz.  A further splitting of each HF state results in a
\emph{superhyperfine (SHF)} structure ($\nu_{\mathrm{SHF}} = 150$-300~MHz),
caused by the interaction of the $\pbar$ spin $S_{\pbar}$ with $F = L +
S_{\mathrm{e}}$.  There exists therefore a quadruplet substructure for each
($n,\,l$) state as shown in figure~\ref{fig:subfig:HFS}.  The theoretical
framework for
the level splitting has been developed by Bakalov and Korobov~\cite{Bakalov:98}.

\begin{figure}
\subfloat[]{
\label{fig:subfig:trans}
\includegraphics[scale=0.7]{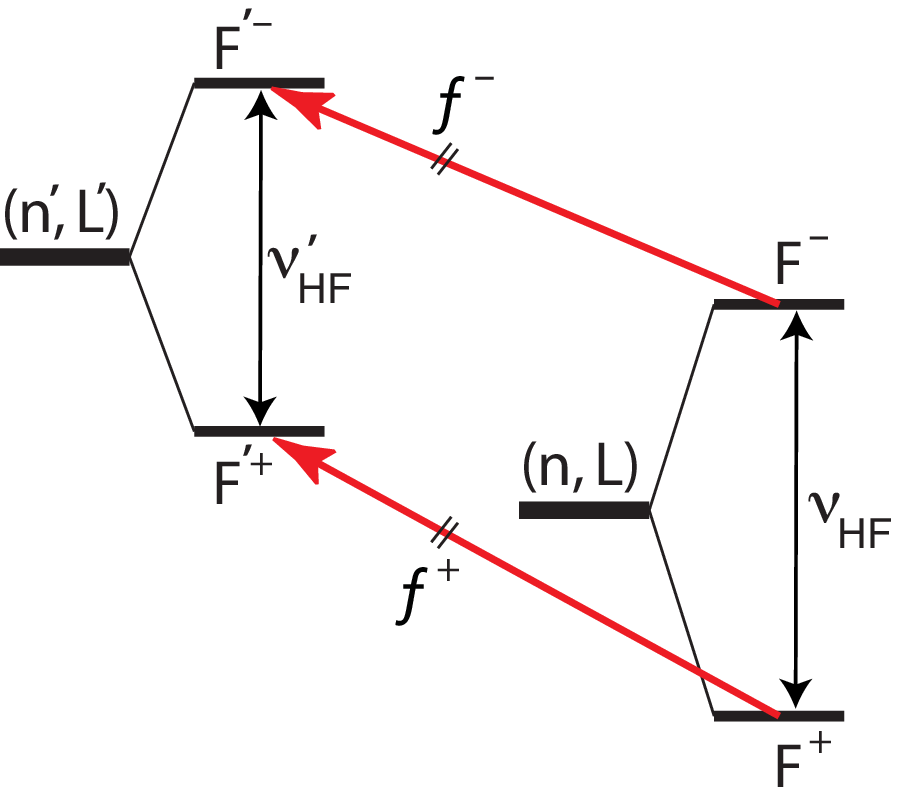}}
\subfloat[]{
\label{fig:subfig:laser}
\includegraphics[scale=0.55]{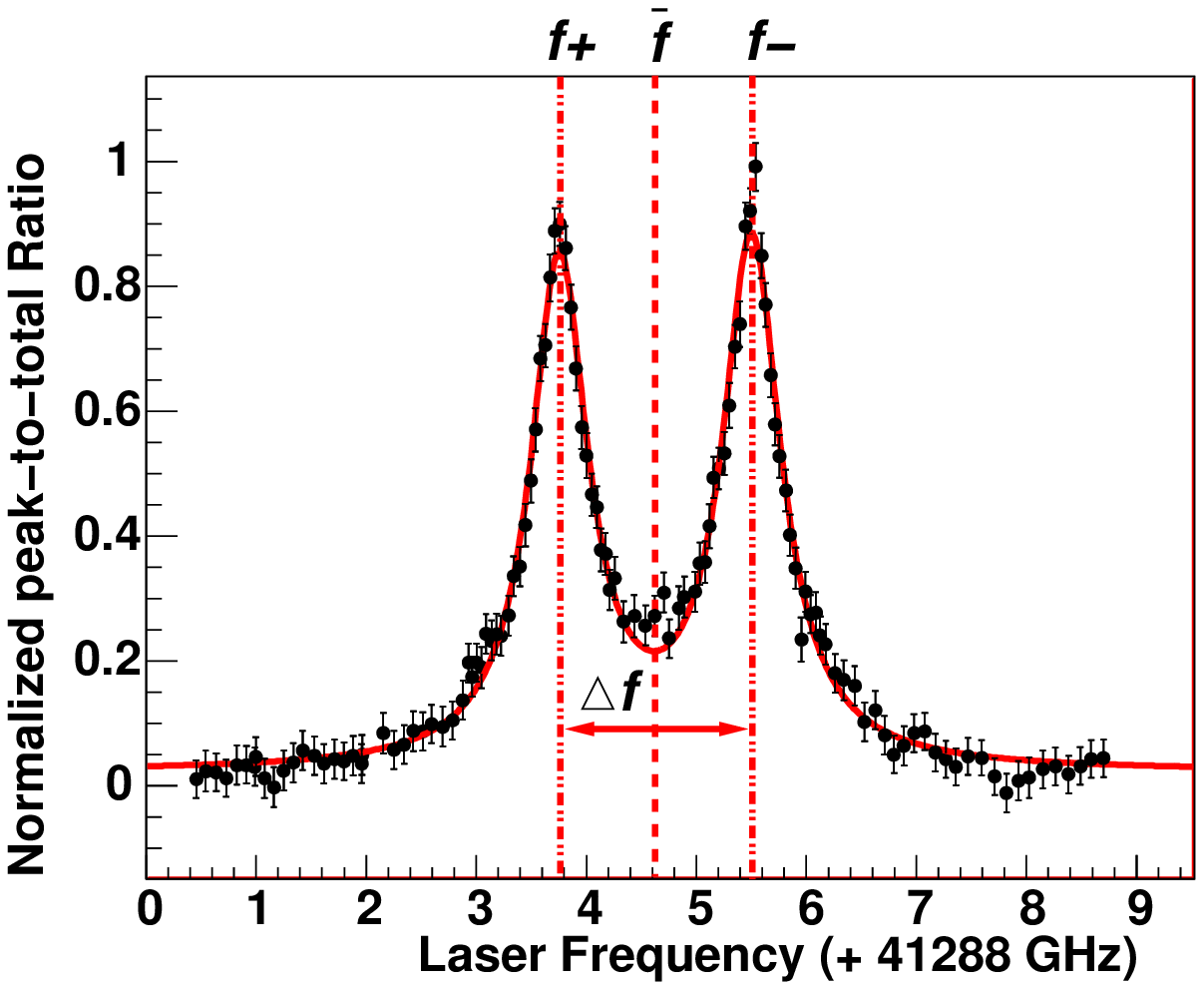}}
\caption{(a) Schematic view of the primary level splitting of $\pbHe$ for the
unfavoured electric dipole transitions.  The state drawn on the right is the
radiative decay dominated parent $(n, \, L)$, and  state on the left is the
Auger
decay dominated daughter $(n', \, L')$. The laser transitions, from the parent
to the daughter doublets, are indicated by the arrows $f^+$ and $f^-$. (b) Laser
resonance profile of the $(n,l) = (37, \, 35)$ to $(38, \, 34)$ transition
showing the $f^+$ and $f^-$ transitions, from~\cite{Pask:2008}}
\label{fig:trans}
\end{figure}

The HF doublet is described by the quantum number $\vec{F} = \vec{L} +
\vec{S}_{e}$ with components $F^{+} = L + \frac{1}{2}$ and $F^{-} = L -
\frac{1}{2}$.  The SHF quadruplet is described by $\vec{J} = \vec{F} +
\vec{S}_{\pbar}$ with components $J^{-+} = F^{-} + \frac{1}{2}$, $J^{--} = F^{-}
- \frac{1}{2}$, $J^{++} = F^{+} + \frac{1}{2}$ and $J^{+-} = F^{+} -
\frac{1}{2}$.  Between these sub states an electron spin flip can be induced by
two M1 transitions $\nu^{+}_{\mathrm{HF}}$ and $\nu^{-}_{\mathrm{HF}}$ (see
figure~\ref{fig:subfig:HFS}):

\numparts
\begin{eqnarray}
\nu^{+}_{\mathrm{HF}}: \quad J^{++} = F^{+} + \frac{1}{2} = L + 1 &
\leftrightarrow & J^{-+} = F^{-} + \frac{1}{2} = L, \\
\nu^{-}_{\mathrm{HF}}: \quad J^{+-} = F^{+} - \frac{1}{2} = L & \leftrightarrow
& J^{--} = F^{-} - \frac{1}{2} = L - 1.
\end{eqnarray}
\endnumparts

Electric dipole transitions (E1) between different levels of the cascade can be
induced with resonant laser light\cite{Torii:99,Hori:01,Hori:03,Hori:06}.  There
are two types: favoured, $\Delta v = 0$; $(n,l) \rightarrow (n - 1, l - 1)$, and
 unfavoured, $\Delta v = 2$; $(n, \, l) \rightarrow (n + 1, \, l -
1)$~\cite{Yamazaki:02}.  The dipole moment of the unfavoured transitions is an
order
of magnitude smaller than that of the favoured.  To the first order, atoms
occupying the $F^-$ doublet of the radiative decay dominated  state are
transferred to the $F^{\prime-}$ doublet of the Auger dominated state, while
those occupying $F^+$ are transferred to $F^{\prime +}$, shown in
figure~\ref{fig:subfig:trans}. These transitions are labelled $f^+$ and $f^-$
respectively and the difference between them $\Delta f$.  The unfavoured
transitions have $\Delta f = 1.5-1.8$~GHz, while the favoured have $\Delta f
\leq 0.5$~GHz.  The Doppler broadening at the target temperature is $\sim
0.3-0.5$~GHz, therefore only the unfavoured HF laser transitions can be well
resolved, see figure~\ref{fig:subfig:laser}.

\section{Motivation and Method}

A precise measurement of the $\pbHe$ HF
splitting~\cite{Widmann:97,Widmann:02,Pask:2008} is of great importance for
rigorously testing three-body quantum electrodynamic (QED)
calculations~\cite{Bakalov:98,Korobov:01,Yamanaka:01,Kino:03APAC}, leading to a
determination of the antiproton spin magnetic moment and a test of \emph{CPT}
invariance~\cite{Pask:2009}.  For an understanding of the collision processes
between $\pbHe$ atoms and the He atoms of the medium, a comparison between
experiment and theory can be equally useful.  Both the elastic $\Gamma_{e}$ and
inelastic $\Gamma_{i}$ collision rates can have significant systematic effects
on experimental results.  Elastic collisions contribute to a shift and a
broadening $\Delta - i\Gamma_{e}$, while inelastic collisions result in a spin
exchange, therefore a change of state.

A clear understanding of the collision processes was essential for the
interpretation of the E1 spectral lines~\cite{Hori:02}, a similar study has been
performed for the M1 transitions.  To measure $\Delta - i \Gamma_{e}$, 
microwave
resonant profiles were scanned at various target gas densities.  The line width
is limited by the Fourier transform of the microwave pulse length but
observation of a larger width would be evidence of a collisional broadening. 
Likewise a density dependent change in the transition frequencies would be
evidence of a collisional shift.

The results of elastic collisional studies have been presented in previous
publications~\cite{Pask:2008,Pask:2009p} and indicate that $\Gamma_{e}$ is small
because the dominating broadening effect is found to be from the Fourier
transform of the microwave pulse length.  Korenman predicts that $\Gamma_{e}
\sim 2.5\Gamma_{i}$~\cite{Korenman:06} which means that if $\Gamma_{e}$ is
smaller than first predicted then so must $\Gamma_{i}$.

The inelastic collision rate was determined by measuring the
time dependence of the $F^+$ population. Two narrow-band lasers were tuned to
the $f^{+}$ transition between the radiative decay dominated parent state $(n,
\, L) = (37, \, 35)$ and the Auger decay dominated daughter state $(38, \, 34)$
shown in figure~\ref{fig:trans}.  The second was delayed by a time $T =
50-2000$~ns from the first.

\begin{figure}
\begin{center}
\includegraphics[scale=0.5]{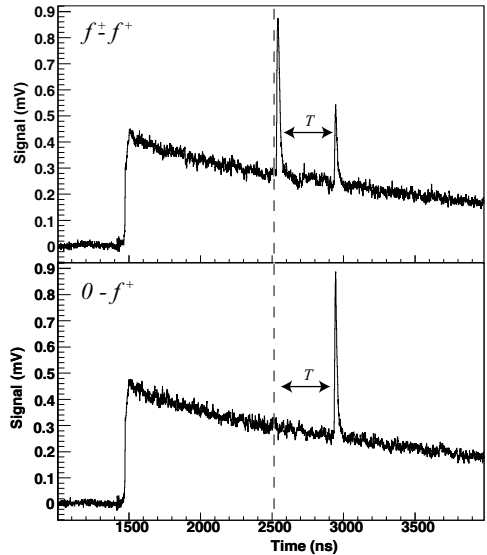}
\caption{Laser stimulated annihilation peaks against the exponential decaying
background of the other metastable states' populations.  Mode 1):
$f^{+}$-$f^{+}$ when both lasers are fired and mode 2): $0$-$f^{+}$ when only
the second laser is fired.  The peak in Mode 2) is larger than the second peak
in Mode 1) because no previous depopulation of the state has been induced.}
\label{fig:AA_OA}
\end{center}
\end{figure}

The $\pbar$ annihilation products were detected with Cherenkov counters as a
function of time.  The metastable tail, where the radiative decay dominated
states cascade towards the nucleus, was recorded as background.  Because of its
short lifetime ($\sim 10$~ns), the laser resonant transfer to an Auger dominated
decay state results in a sharp peak in annihilations events which stands out
against
the background, shown in figure~\ref{fig:AA_OA}.  The ratio between this peak
area to the area under the entire spectrum (peak-to-total) is proportional to
the population transferred with the laser.  The peak-to-total of the first and
second laser annihilation peaks are represented by $r_{1}$ and $r_{2}$,
respectively.

The experiment was performed in two different modes: 1) where both lasers were
fired, $f^{+}$-$f^{+}$ and 2) where only the second laser was fired,
$0$-$f^{+}$,
both of which are shown in figure~\ref{fig:AA_OA}.  Mode 1) contained all the
information about $\Gamma_i$ while mode 2) was required to extract information
about the refilling from higher states, also contained in 1).

The method employed in \cite{Widmann:02}, \cite{Pask:2008} and \cite{Pask:2009}
to
determine the HF splitting included a microwave pulse between the first and
second lasers of mode 1).  By scanning the microwave over a range of frequencies
and measuring the dependence of $r_{2}$, the $\nu_{\mathrm{HF}}^+$ and
$\nu_{\mathrm{HF}}^-$ resonances were found.  The maximum achievable signal
for a range of laser delays was determined by fixing the microwave frequency to
one
transition, say $\nu_{\mathrm{HF}}^+$, and monitoring $r_{2}$ while scanning the
microwave power.

\section{Apparatus}\label{sec:App}

The experiment was performed at CERN's Antiproton Decelerator (AD), which
delivered a pulse of 1-$4 \times 10^7$ antiprotons with a length of 200~ns
(FWHM) and an energy $E = 5.3$~MeV at $\sim 90$~s intervals.  Antiprotonic
helium
was formed by stopping antiprotons in a gas target at a temperature of 6.1~K and
a pressure $p = 150$-500~mbar (number density 1.7-$6.2 \times
10^{20}$~cm$^{-3}$).

Charged pions were produced by antiproton annihilations in the helium nucleus
and could be detected by Cherenkov counters.  The signal was amplified by
fine-mesh photomultipliers (PMTs) and the resulting analog delayed annihilation
time spectrum (ADATS) was recorded in a digital oscilloscope (DSO). The PMTs
were gated off during the $\pbar$ pulse arrival so that only the 3\% metastable
tail was recorded~\cite{CherHori:03}.

Two pulse-amplified laser beams were produced by splitting a continuous wave
(cw) laser beam of wave-length 726.1~nm into two seed beams~\cite{Pask:2008}. 
These were pulsed by amplifying the seeds using dye filled Bethune cells pumped
by two pulsed Nd:Yag lasers, the second delayed by time $T$ after the first. 
The pump beams were stretched so that the two pulse lengths were of the order
$\sim 15$~ns~\cite{Hori:06} and therefore similar to the Auger decay rate
ensuring a high depopulation efficiency.  The maximum emitted energy fluence at
the target was $\sim 30$~mJ/cm$^2$ with a spot diameter of 5~mm.

To measure the HF transitions, a
microwave pulse was produced by a vector network analyzer (Anritsu
37225B) referenced to a 10~MHz GPS (HP~58503B) satellite signal and amplified by
a pulsed travelling wave tube amplifier (TMD PTC6358).  A cylindrical resonant
microwave cavity with central frequency $\nu_0 = 12.91$~GHz provided the desired
shape for the field (TM$_{110}$ mode) at the target.  To cover the $\Delta \nu 
\sim 100$~MHz microwave scanning range, the cavity was over-coupled to the wave
guide so that its loaded quality factor $Q_L$ was $\sim 100$, where $\Delta \nu
= f_0 / Q_L$~\cite{Sakaguchi:04-NIM}.  Most of the power was reflected back
towards the amplifier and absorbed by an isolator.  An antenna was connected to
the cavity to monitor the field so that the desired power could be achieved by
controlling the amplification of the pulse.

\section{Analysis}

\subsection{Mode 1) ($f^{+}$-$f^{+}$)}

\begin{figure}[tb]
\begin{center}
\includegraphics[scale=0.5]{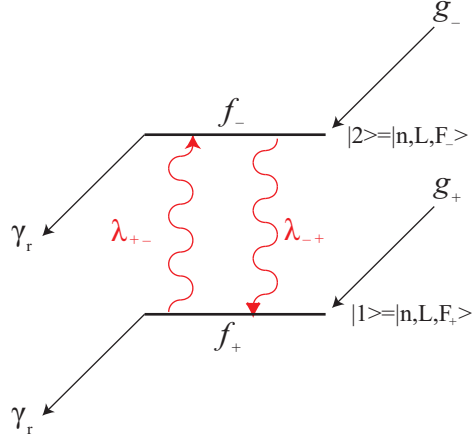}
\caption[Energy level diagram of part of the cascade]{Energy level diagram of
part of the cascade showing the refilling $g_{\pm}(t)$ from above states, decay
to lower states $\gamma_r$ and the relaxation collision rate $\lambda_{\pm
\rightarrow \mp}$.} \label{fig:ccascade}
\end{center}
\end{figure}

When both lasers were fired, the second delayed by a period $T$ after the first,
the normalized peak-to-total $r_{2}/r_{1}$ was plotted as a function of $T$. 
The data were fitted with a function derived from the integral of the following
two equations:

\numparts
\begin{equation}\label{eq:chloe1}
\mathrm{\frac{d\rho_+}{dt}} = g_+(t) - (\lambda_{+-} + \gamma_r)\rho_+ +
\lambda_{- +}\rho_-,
\end{equation}

\begin{equation}\label{eq:chloe2}
\mathrm{\frac{d\rho_-}{dt}} = g_-(t) - (\lambda_{-+} + \gamma_r)\rho_- +
\lambda_{+-}\rho_+,
\end{equation}
\endnumparts

\noindent where $\rho_{\pm}$ is the population density of the HF states, and
$g_{\pm}(t)$ describes the refilling rate from the higher lying states.   The
relaxation rates from $\rho_+$ to $\rho_-$ and $\rho_-$ to $\rho_+$ are
represented by $\lambda_{+-}$ and $\lambda_{-+}$ respectively.  The radiative
decay rate is $\gamma_r = 7.149 \times 10^5$~s$^{-1}$~\cite{Yamazaki:02}, see
figure~\ref{fig:ccascade}.

If $t = 0$ is the time when the first laser is fired then the relative 
population of the two levels at $t < 0$ is 

\begin{equation}
\rho_{\pm} = \frac{2 F^{\pm} + 1}{2(2 L + 1)},
\end{equation}

\noindent where $F^+ = L+\frac{1}{2}$ and $F^-=L-\frac{1}{2}$.  The signal from
the first and second laser are therefore
\numparts

\begin{equation}\label{eq:ptt1}
r_1 = I_0 \frac{L + 1}{2L + 1} \epsilon_1,
\end{equation}

\begin{equation}\label{eq:ptt2}
r_2(t) = I_0 \rho_+(t) \epsilon_2,
\end{equation}
\endnumparts

\noindent where $I_0$ is a normalization factor and $\epsilon_{1}$ and
$\epsilon_{2}$ are the laser depopulation efficiencies for the first and second
laser.  The overlap of the laser peaks, whereby the $F^{-}$ transition is
partially induced due to the Doppler broadening when the laser is tuned to the
$f^+$ transition, is considered negligible in this analysis.

At $t = 0$ and since $L \gg 1$, the initial populations are

\numparts
\begin{equation}
\rho_+(0) = \frac{L + 1}{2L + 1} (1 - \epsilon_1) \approx  \frac{1}{2}(1 -
\epsilon_1),
\end{equation}

\begin{equation}
\rho_-(0) = \frac{L}{2L + 1} \approx \frac{1}{2}.
\end{equation}
\endnumparts

\noindent It can also be assumed that $\lambda_{+-}=\lambda_{-+} \equiv
\Gamma_{i}$ and $g_{+}=g_{-} \equiv g$.  So (\ref{eq:chloe1}) and
(\ref{eq:chloe2}) can be written as follows

\numparts
\begin{equation}\label{eq:simp1}
\rho_+(t) = \frac{1}{2} \mathrm{e}^{-\gamma_r t}\Big(1 - \frac{\epsilon_1}{2} +
\frac{\epsilon_1}{2} \mathrm{e}^{-2 \Gamma_{i} t}\Big) + \frac{1}{2}\int^{t}_{0}
\mathrm{e}^{-\gamma_r (t-t^{\prime})} \mu_f  h(t^{\prime}) \mathrm{d}t^{\prime},
\end{equation}

\begin{equation}\label{eq:simp2}
\rho_-(t) = \frac{1}{2} \mathrm{e}^{-\gamma_r t}\Big(1 - \frac{\epsilon_1}{2} -
\frac{\epsilon_1}{2} \mathrm{e}^{-2 \Gamma_{i} t}\Big) + \frac{1}{2}
\int^{t}_{0} \mathrm{e}^{-\gamma_r (t-t^{\prime})} \mu_f  h(t^{\prime})
\mathrm{d}t^{\prime},
\end{equation}
\endnumparts

\noindent where $g = \mu_f h(t)$, of which $\mu_f$ is a constant
associated with the rate of filling from above states and $h(t)$ is the
filling function normalized by $h(0) = 1$.  The simplest assumption for
$h(t)$ is $h(t) = \mathrm{e}^{-\mu_0 t}$ where
$\mu_0$ is a decay rate associated with the population of the above states. This
simplification was necessary to achieve an unambiguous
result due to the limited amount of data. So the solution to the integral in
(\ref{eq:simp1}) and (\ref{eq:simp2}) becomes
 
\begin{equation}
F(t) \equiv \mu_f \int^{t}_{0} \mathrm{e}^{-\gamma_r (t-t^{\prime})}
h(t^{\prime}) \mathrm{d}t^{\prime} =  \frac{\mu_f  (\mathrm{e}^{-\gamma_r t} -
\mathrm{e}^{-\mu_0 t})}{\mu_0 - \gamma_r}.
\end{equation}

\noindent At $T < 1000$~ns the filling can be assumed constant $\mu_0 = 0$, and
thus

\begin{equation}\label{eq:const}
F(t) =  \frac{\mu_f ( 1 - \mathrm{e}^{-\gamma_r t})}{\gamma_r}.
\end{equation}

\noindent Substituting (\ref{eq:const}) into (\ref{eq:simp1}), then
(\ref{eq:simp1}) into (\ref{eq:ptt2}) and normalizing
over (\ref{eq:ptt1}) gives:

\begin{equation}\label{eq:AA}
\frac{r_2(t)}{r_1(t)} = \frac{\epsilon_2}{\epsilon_1}\Big[e^{-\gamma_{r}t} \Big(
1 - \frac{\epsilon_1}{2} - \frac{\epsilon_1}{2} e^{-2 \Gamma_{i} t} -
\frac{\mu_f}{\gamma_{r}} \Big) + \frac{\mu_f}{\gamma_{r}}\Big].
\end{equation}

\subsection{Mode 2) ($0$-$f^{+}$)}

Only firing the second laser, the population decay and refilling of the state
can
be measured.  In this regime $r_{2}$ was normalized to the average $r_{1}$ from
mode 1).  This was performed so that both sets of data could be plotted on the
same scale and compared adjacently.    There are no inelastic collision
terms because no asymmetry is created,

\begin{equation}\label{eq:OA}
\frac{r_{2}(t)}{r_{1}(t)} = \frac{\epsilon_2}{\epsilon_1}\Big[e^{-\gamma_{r}t}
\Big( 1 -  \frac{\mu_f}{\gamma_{r}} \Big) + \frac{\mu_f}{\gamma_{r}}\Big].
\end{equation}

\subsection{Numerical simulation}\label{sec:NumSim}

The $(37, \, 35)\rightarrow (38, \,34)$ laser transition was numerically
simulated by evolving the optical Bloch equations, obtaining a maximum
depopulation efficiency of 70\%.  

The microwave transitions between the HF
substates were determined by evolving (\ref{eq:trans4}), derived from the
optical Bloch equations, which can be written as two independent $4 \times 4$
matrices to handle the $\nu_{\mathrm{HF}}^+$ and $\nu_{\mathrm{HF}}^-$
transitions separately.  However, collision induced $\pbar$ spin flips  result
in $\nu_{\mathrm{SHF}}^+$ and $\nu_{\mathrm{SHF}}^-$ transitions.  Thus the 
population evolutions of the $J^{-+}$ and $J^{++}$ states become dependent on
those of the $J^{--}$ and $J^{-+}$ states.  The resulting simultaneous equation
has the solution of the form of an $8 \times 8$ matrix.   An additional two
dimensions were added to simulate the refilling from above states:

\numparts
\begin{equation}\label{eq:trans4}
\mathrm{\frac{d}{dt}} \boldsymbol{\rho} = \mathbf{M} \boldsymbol{\rho},
\end{equation}

{\fontsize{10}{8}
\begin{equation}
\boldsymbol{\rho} = \left( \begin{array}{c}
\rho_{-+} \\
\rho_{++} \\
\rho_{x+} \\
\rho_{y+} \\
\rho_{--} \\
\rho_{+-} \\
\rho_{x-} \\
\rho_{y-} \\
\rho_{u38} \\
\rho_{u39} \\
\end{array} \right),
\end{equation}

\begin{equation}\fl
\mathbf{M} = \left( \begin{array}{cccccccccc} -\gamma_{c} & \Gamma_{i} & 0 &
\frac{1}{2}\Omega^+_m & \Gamma_{i} & 0 & 0 & 0 & \gamma_{u38}/4 & 0\\
\Gamma_{i} & -\gamma_{c} & 0 & -\frac{1}{2}\Omega^+_m & 0 & \Gamma_{i} & 0 & 0 &
\gamma_{u38}/4 & 0 \\
0 & 0 & -\gamma_{T} & \Delta\omega_+ & 0 & 0 & 0 & 0 & 0 & 0\\
-\Omega^+_m & \Omega^+_m & -\Delta\omega_+ & -\gamma_{T} & 0 & 0 & 0 & 0 & 0 &
0\\

\Gamma_{i} & 0 & 0 & 0 & -\gamma_{c} & \Gamma_{i} & 0 &
\frac{1}{2}\Omega^-_m & \gamma_{u38}/4 & 0\\ 0 & \Gamma_{i} & 0 & 0 &
\Gamma_{i} & -\gamma_{c} & 0 & -\frac{1}{2}\Omega^-_m &  \gamma_{u38}/4 & 0 \\
0 & 0 & 0 & 0 & 0 & 0 & -\gamma_{T} & \Delta\omega_- & 0 & 0\\ 0 & 0 & 0 & 0 &
-\Omega^-_m & \Omega^-_m & -\Delta\omega_- & -\gamma_{T} & 0 & 0\\

0 & 0 & 0 & 0 & 0 & 0 & 0 & 0 & -\gamma_{u38} & \gamma_{u39}\\
0 & 0 & 0 & 0 & 0 & 0 & 0 & 0 & 0 & -\gamma_{u39}\\
\end{array} \right),
\end{equation}
}
\endnumparts

\noindent where $\rho_{-+}, \rho_{++}, \rho_{--}$, and $\rho_{+-}$ represent the
different time dependent populations of the four SHF states of the $(n,
l) = (37, \, 35)$ state.  In a field free region $\Omega_m = 0$, these
populations
simply decay radiatively  at a rate of $\gamma_{r} = 7.149 \times 10^5$~s$^{-1}$
to the ($36, \, 34$) state and $0.0086 \times 10^5$~s$^{-1}$ to the ($37, \, 34$) state
\cite{Yamazaki:02}.  When there is a population asymmetry and an external
oscillating magnetic field is present, transfer between the states can be
observed.  The complex dependency of the transitions is represented by
$\rho_{x\pm}$ and $\rho_{y\pm}$ for the real and imaginary parts respectively.
Broadening effects are dependent on both the radiative decay rate  $\gamma_r$
and the elastic  collisional frequency $\Gamma_{e}$: $\gamma_T = \gamma_r +
\Gamma_{e}$.

The four SHF states are refilled as the upper states, (38,36) $\rho_{u38}$ and
(39,37) $\rho_{u39}$, decay at a rate of $\gamma_{u38} = 6.55 \times
10^5$~s$^{-1}$ and $\gamma_{u39} = 5.88 \times 10^5$~s$^{-1}$ into the lower
($37, \, 35$) state, as part of the cascade \cite{Yamazaki:02}. The initial
populations of these states have been experimentally measured
\cite{Horip:04,Hori:02}.  Through inelastic relaxation collisions $\Gamma_{i}$
the atoms return to an equilibrium, the variable $\gamma_{c}$ is defined as
$\gamma_{c} = 2 \Gamma_{i} + \gamma_{r}$.
 Collisions which result in the spin flip of more than one particle are ignored
\cite{Korenman:06}.

The offset between the microwave frequency $\nu_{\mathrm{M}}$ and transition
 frequencies $\nu_{\mathrm{HF}}$ is represented by $\Delta \omega = 2
\pi$($\nu_{\mathrm{M}} - \nu_{\mathrm{HF} \pm}$).  The Rabi frequency is
dependent upon the magnetic field strength $B$ and the atom's magnetic dipole
moment $\mu_m$

\begin{equation}\label{eq:rabimic}
\Omega_m = \frac{\mu_m B(x,y,t)}{\hbar},
\end{equation}
\begin{equation}
\mu_m = \langle n',L',F',J',m| \mu_M |n,L,F,J,m \rangle,
\end{equation}

\noindent which can be calculated using Wigner's 3-j and Racah's 6-j
coefficients


\begin{figure}[tb]
\begin{center}
\includegraphics[scale=0.4]{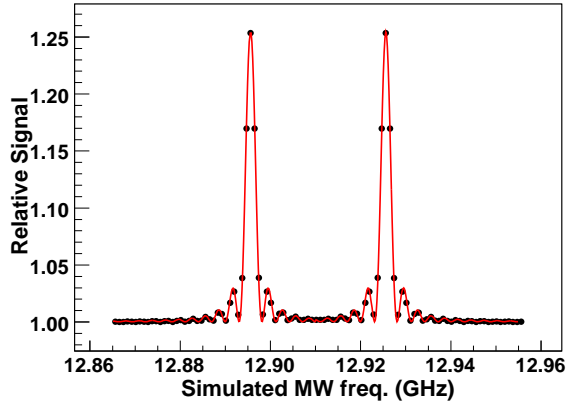}
\caption{Simulated microwave resonance profile, where $T = 350$~ns $\Gamma_i =
3.4 \times 10^5$~s$^{-1}$, fitted with the same function as the experimental
measurements presented in \cite{Pask:2008} and
\cite{Pask:2009}}\label{fig:SimScan}
\end{center}
\end{figure}

\begin{displaymath}
\langle n,L,F^-,J',m| \mu |n,L,F^+,J,m \rangle = (-1)^{J'+
m}\left( \begin{array}{ccc} J & 1 & J'\\
m & 0 & -m \end{array} \right)_{3-j}
\end{displaymath}
\begin{equation}\label{eq:rabimic2}
\times \sqrt{(2J+1)(2J'+1)} \left\{
\begin{array}{ccc} F^- & J' & \frac{1}{2} \\
J & F^+ & 1 \end{array} \right\} _{6-j} \quad \quad \quad
\quad \quad \quad \quad \quad \quad
\end{equation}
\begin{displaymath}
\times \sqrt{(2F^+ +1)(2F^-+1)} \left\{
\begin{array}{ccc} \frac{1}{2} & F^- &  L\\
F^+ & \frac{1}{2} & 1 \end{array} \right\} _{6-j}
g_{\mathrm{e}}\mu_b\langle \frac{1}{2}| s_{\mathrm{e}} | \frac{1}{2} \rangle
\quad \quad
\quad \quad \quad,
\end{displaymath}

\noindent where $n, \, L, \, F, \, J, \, m$ are the corresponding quantum
numbers.

Equation~(\ref{eq:trans4}) was solved over a range of 60 equally spaced
frequencies $\nu_{\mathrm{M}}$ to simulate a microwave resonance profile
measurement.  The population positional
distribution and magnetic field variance were modelled with a Monte Carlo
positional simulation.  The magnetic field distribution at the target is
dependent on the cavity and  varies spatially with respect to the radial $r$ and
angular cylindrical $\phi$ co-ordinates \cite{Sakaguchi:04-NIM}. Apart from edge
effects the  cylindrical component $z$ is constant. The radial $B_r$ and angular
$B_\phi$ components of the magnetic field are given by

\numparts
\begin{equation}
B_r(r,\phi)=B_0\frac{J_1(kr)}{kr}\sin(\phi),
\end{equation}

\begin{equation}
B_\phi(r,\phi)=B_0J'_1(kr)\cos(\phi),
\end{equation}
\endnumparts

\noindent where $k$ is the wave number and $J_1$ is the Bessel function of the
first kind.  The stopping distribution of the $\pbHe$ is assumed to be Gaussian
in both the $z$ and $r$ planes~\cite{Sakaguchi:04-NIM}.

\section{Results}

\begin{figure}[tb]
\begin{center}
\includegraphics[scale=0.4]{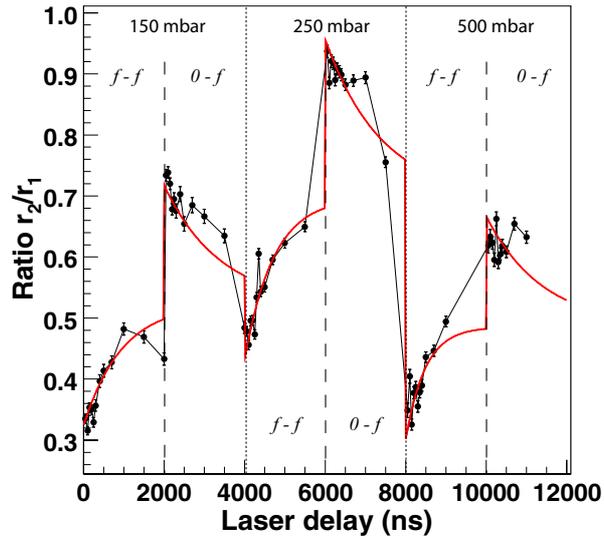}
\caption{Simultaneous fitting of the data for $p = 150$~mbar, 250~mbar and
500~mbar.  Mode 1) ($f^+$-$f^+$) was fitted with~(\ref{eq:AA}) and mode 2)
(0-$f^+$) was fitted with~(\ref{eq:OA}).  The refilling rate $\mu_f$ was a
common parameter for all target densities and the laser depopulation
efficiencies $\epsilon_1$ and $\epsilon_2$ were common parameters for different
modes measured at the same density.}\label{fig:SimFit}
\end{center}
\end{figure}

Data were measured at three different target pressures $p = 150$~mbar, 250~mbar
and 500~mbar resulting in a total of six graphs; three for mode 1) and three
for mode 2).  These were plotted side by side and fitted simultaneously
with (\ref{eq:AA}) and (\ref{eq:OA}).  The variables $\mu_f$, $\epsilon_1$ and
$\Gamma_i$ were common for all pressures, where the latter was weighted
proportionally to the target gas density and $\epsilon_2$ was left free for
different target densities.  Other, more complex fit functions were also
attempted.  These varied to include the population evolution of the upper levels
and left $\epsilon_1$ free for different target densities.  However, the
introduction of more parameters limited convergence and put emphasis on the
refilling processes.  It was found that the simplest function provided the most
sensitivity to $\Gamma_i$.

\begin{figure}[tb]
\begin{center}
\includegraphics[scale=0.4]{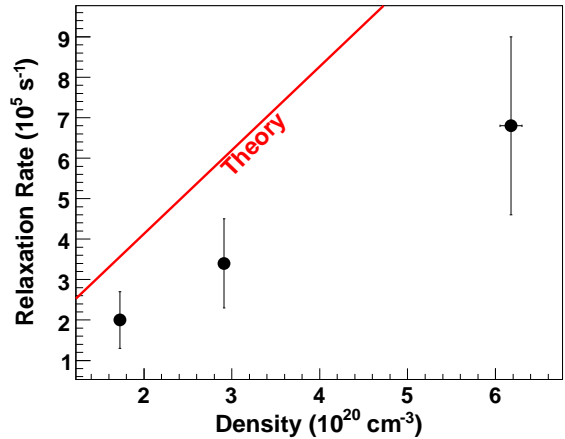}
\caption{Dependence of the collision induced relaxation rate $\Gamma_{i}$ on the
helium target density.} \label{fig:RelRate}
\end{center}
\end{figure}

A graph of the data fitted with~(\ref{eq:AA}) and~(\ref{eq:OA}) is shown in
figure~\ref{fig:SimFit}.  The collision induced relaxation rate has been
determined from the fitting parameters and plotted in figure~\ref{fig:RelRate}. 
The numerical values are displayed in table~\ref{tab:RelRate} except for $\mu_f$
and $\epsilon_1$ which were determined to be $(5.2 \pm 0.2) \times
10^5$~s$^{-1}$  and 55\% respectively.

\begin{table}[h]
\begin{center}
\caption{}\label{tab:RelRate}
\begin{tabular}{ c | c | c | c | c | c}
\br
$p$~(mbar) & $\rho_{\mathrm{He}}$ (10$^{20}$~cm$^{-3}$) & $\epsilon_{2}$ (\%) &
$\Gamma_{i} (10^5$~s$^{-1}$) & $\tau_{i}$~(ns) \\

\mr
150 & 1.726 & 39(2) & 2.0(0.7) & 2500(900) \\
250 & 2.912 & 53(2) & 3.4(1.1) & 1500(500) \\
500 & 6.177 & 37(3) & 6.8(2.2) & 750(250) \\
\br
\end{tabular}
\end{center}
\end{table}

The errors associated with these relaxation rates have been inflated by the
square root of the reduced chi squared $\chi^2_{red} \sim 10$ of the fit but
still remain 2-3 sigma less than the most recent theoretical calculations which
predict $\Gamma_{i} = 6.2 \times 10^5$~s$^{-1}$~\cite{Korenman:2009p} for $p =
250$~mbar.  The laser depopulation efficiency is also revealed to be smaller
than the 70\% predicted in Section \ref{sec:NumSim}.  The fluctuations of
$\epsilon_2$, that can vary by as much as 15\%, are put down to the fact
that data of different densities were measured on different days.  The
fluctuations are  therefore most probably caused by changes in the overlap of
the two lasers with the $\pbar$ beam and each other.

\begin{figure}[tb]
\begin{center}
\includegraphics[scale=0.4]{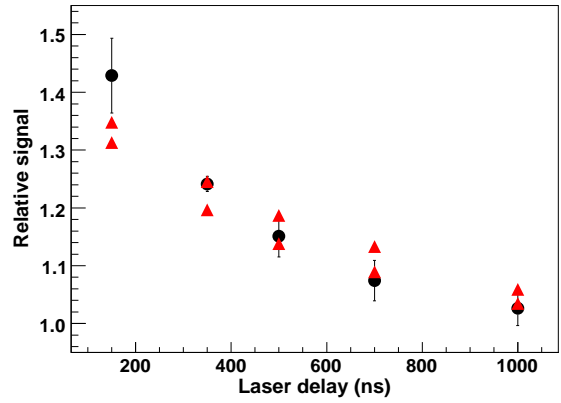}
\caption{The optimum signal-to-noise ratio for each time delay $T$ at
$p = 250$~mbar.  The circles ({\LARGE $\bullet$}) represent the experimental
results from Pask \emph{et.~al.}~\cite{Pask:2008} while the triangles
(\textcolor{red}{$\blacktriangle$}) represent the simulated data.  The larger
signal results from $\Gamma_{i} = 2.3 \times 10^5$~s$^{-1}$ and the smaller from
$\Gamma_{i} = 4.5 \times 10^5$~s$^{-1}$} \label{fig:RelComp}
\end{center}
\end{figure}

To be certain that the fit provided a realistic determination of the collision
parameter, microwave resonant scans were simulated as described in Section
\ref{sec:NumSim}.  The simulated signal-to-noise ratio for $T = 150$~ns, 350~ns,
500~ns, 700~ns and 1000~ns was compared to experimental data measured at
$p = 250$~mbar from~\cite{Pask:2008}.  It was assumed that $\Gamma_{e}=2.5
\Gamma_{i}$~\cite{Korenman:06} and $\epsilon_{1} = \epsilon_{2} = 55\%$. The
upper and lower limits, determined from the 1 sigma error associated with
$\Gamma_{i}$, are plotted alongside the previously measured data, shown in
figure~\ref{fig:RelComp}.

At small $T$, $\Gamma_{i} = 4.5 \times 10^5$~s$^{-1}$ tends to underestimate the
signal-to-noise ratio while $\Gamma_{i} = 2.3 \times 10^5$~s$^{-1}$
overestimates the signal when $T$ is large.  At $T = 200$~ns both the upper and
lower limits are too small.  This is due to the laser depopulation efficiency,
which has a larger effect on the signal at short $T$.  The experimental data
displayed in figure~\ref{fig:RelComp} were measured during a different year
(2006) to those data displayed in figure~\ref{fig:RelRate} (2008).  During this
time the $\pbar$ flux was larger with higher stability and therefore
$\epsilon_{1}$ and $\epsilon_{2}$ were likely to be higher than during 2008.

\section{Conclusions}

This study of the collision rates between $\pbHe$ atoms and the He medium has
been used both to determine optimal conditions for microwave resonance profile
measurements and to compare with theoretical predictions so that the system may
be better understood.  Experimental measurements of elastic collisions published
in previous papers~\cite{Pask:2009p} prompted a re-evaluation of
theory~\cite{Korenman:06} which had originally overestimated the collision rate.
 Since the cross-sections of the two collision processes were predicted to be
similar, a measurement of the
inelastic collision rate was expected to introduce valuable knowledge about the
interactions.

For the first time inelastic spin exchange collisions have been been measured
between the HF states of $\pbHe$.  Given the complexity of the system and the
uncertainty in determining the initial parameters of the theoretical model, the
measured values are in agreement with the recent theory~\cite{Korenman:2009p}. 
More rigorous calculations are anticipated for a more thorough comparison.

The laser depopulation efficiency is shown to be smaller than predicted but also
to
depend heavily on the conditions of the $\pbar$ beam and alignment.  Simulations
comparing the results to earlier data indicate agreement between the two methods
but the inelastic collision rate is most likely to tend towards theory in the
upper limits of experimental uncertainty.

Collisional effects in E1 transitions have previously been shown to vary
depending on the state measured, therefore other states are of interest. A study
of the collision processes in the $\pbar ^3$He$^+$ HF structure, which contains
an additional degree of freedom due to the helion spin, is also planned for
future work.

\ack

The authors would like to acknowledge many fruitful discussions with
Prof.~Grigory Korenman (Moscow State University).  We thank the AD operators for
providing the antiproton beam.  This work was supported by Monbukagakusho (grant
no.~15002005), by the Hungarian National Research Foundation (NK67974 and
K72172), the EURYI Award of the European Science Foundation and the Deutsche
Forschungsgemeinschaft (DFG), the Munich-Centre for Advanced Photonics (MAP)
Cluster of DFG and by the Austrian Federal Ministry of Science and Research.

\section*{References}
\bibliography{ps205,hbar,EBW-new,DH-new,MH-new,RSH-new}

\end{document}